\documentclass[%
 aip,
 amsmath,amssymb,
longbibliography,
reprint,%
]{revtex4-1}
\usepackage{scalerel}
\usepackage{tikz}
\usetikzlibrary{svg.path,calc,shapes,arrows,positioning,shadows,shadings}
\usepackage[colorlinks=true,linkcolor=blue,citecolor=blue,urlcolor=blue]{hyperref}

\usepackage{graphicx}
\usepackage{dcolumn}
\usepackage{bm}
\usepackage[utf8]{inputenc}
\usepackage[T1]{fontenc}
\usepackage{mathptmx}
\usepackage[ruled,vlined]{algorithm2e}
\usepackage{etoolbox}
\usepackage{chemfig}
\usepackage{braket}
\usepackage{simpler-wick}
\usepackage{bm}
\usepackage{float}
\usepackage[version=4]{mhchem}
\usepackage{booktabs}
\usepackage{calc}
\usepackage{booktabs}       
\usepackage{threeparttable}  
\usepackage{adjustbox}  
\usepackage{rotating}

\usepackage{siunitx}
\usepackage{booktabs}
\usepackage{pgfplots}
\makeatletter
\def\@email#1#2{%
 \endgroup
 \patchcmd{\titleblock@produce}
  {\frontmatter@RRAPformat}
  {\frontmatter@RRAPformat{\produce@RRAP{*#1\href{mailto:#2}{#2}}}\frontmatter@RRAPformat}
  {}{}
}%
\makeatother
\include{ccdiag}

\definecolor{orcidlogocol}{HTML}{A6CE39}
\tikzset{
  orcidlogo/.pic={
    \fill[orcidlogocol] svg{M256,128c0,70.7-57.3,128-128,128C57.3,256,0,198.7,0,128C0,57.3,57.3,0,128,0C198.7,0,256,57.3,256,128z};
    \fill[white] svg{M86.3,186.2H70.9V79.1h15.4v48.4V186.2z}
                 svg{M108.9,79.1h41.6c39.6,0,57,28.3,57,53.6c0,27.5-21.5,53.6-56.8,53.6h-41.8V79.1z M124.3,172.4h24.5c34.9,0,42.9-26.5,42.9-39.7c0-21.5-13.7-39.7-43.7-39.7h-23.7V172.4z}
                 svg{M88.7,56.8c0,5.5-4.5,10.1-10.1,10.1c-5.6,0-10.1-4.6-10.1-10.1c0-5.6,4.5-10.1,10.1-10.1C84.2,46.7,88.7,51.3,88.7,56.8z};
  }
}

\newcommand\orcidicon[1]{\href{https://orcid.org/#1}{\mbox{\scalerel*{
\begin{tikzpicture}[yscale=-1,transform shape]
\pic{orcidlogo};
\end{tikzpicture}
}{|}}}}

\usetikzlibrary{shapes.geometric, arrows.meta, positioning, calc, shadows.blur, backgrounds, decorations.markings, fadings}

\definecolor{atom_sr}{RGB}{0, 180, 0}      
\definecolor{atom_o}{RGB}{220, 20, 20}      
\definecolor{atom_s}{RGB}{240, 180, 20}     
\definecolor{tech_dark}{RGB}{30, 40, 60}    
\definecolor{tech_light}{RGB}{240, 245, 255} 
\definecolor{accent_blue}{RGB}{0, 100, 200} 
\definecolor{alert_red}{RGB}{200, 50, 50}   

\begin{document}

\preprint{AIP/123-QED}

\title[]{
 The nuclear electric quadrupole moment of $^{87}$Sr from highly accurate molecular relativistic calculations}
\author{Gabriele Fabbro}
\author{Jan Brandejs}
\author{Trond Saue}\email{trond.saue@irsamc.ups-tlse.fr}
\homepage{https://dirac.ups-tlse.fr/saue}
\affiliation{Laboratoire de Chimie et Physique Quantique,\\UMR 5626 CNRS - Université Toulouse-Paul Sabatier,\\ 118 Route de Narbonne, F-31062 Toulouse, France}

\date{\today}
             
\begin{abstract}
\section*{Abstract}
The nuclear electric quadrupole moment (NQM) of $^{87}$Sr has recently been revisited using high-precision relativistic atomic calculations [B. Lu \textit{et al.}, Phys. Rev. A \textbf{100}, 012504 (2019)], indicating that the currently accepted value should be revised and that their result may serve as a new reference. In the present work, we determine the NQM of $^{87}$Sr from the molecular method, by combining the experimentally measured nuclear quadrupole coupling constants (NQCCs) of SrO and SrS with highly accurate relativistic calculations of the electric field gradient (EFG) at the Sr nucleus. Electronic correlation is treated at the CCSD(T), CCSD-T and CCSD$\tilde{\text{T}}$ levels. The iterative T contribution of the latter, composite scheme was obtained using a newly implemented parallel scheme where the distributed memory tensor library Cyclops Tensor
Framework (CTF) was made available to the DIRAC code for relativistic molecular calculations through TAPP, the new community standard for tensor operations. All correlated calculations are performed using the exact two-component molecular mean-field Hamiltonian (X2C$\mathrm{mmf}$). The Gaunt two-electron interaction is incorporated, an even-tempered optimized quadruple-$\zeta$ quality basis set is employed, and vibrational corrections are accounted for. Our best result is $Q(\ce{^{87}Sr}) = 0.33666 \pm 0.00258$~b, which is about 10\% larger than currently accepted standard value, while it is in excellent agreement with recent determinations [Y.-B. Tang, arXiv:2512.07603 [physics.atom-ph] (2025)].

\end{abstract}

\maketitle

\section{Introduction}
The nuclear electric quadrupole moment (NQM), $eQ$, plays a key role in a wide range of physical phenomena, particularly in the interpretation of hyperfine interactions in spectroscopy.\cite{Neugart_Neyens_2006,Smith1971,Jerschow2005,Pyykkö2018} A non-zero quadrupole moment occurs only for nuclei with spin quantum number $I \ge 1$, manifesting as a deformation of the nuclear charge distribution. Its direct experimental determination is challenging.\cite{Pyykko2008} However, one may use the fact that the NQM interacts with the electric field gradient (EFG) generated by surrounding electrons and other charges (i.e., nuclei). In the principal-axis frame, the EFG is characterized by its largest component, $eq$, and an asymmetry parameter, which vanishes for linear molecules and atoms.\cite{Dean_PhysRev.96.1053} The resulting nuclear quadrupole interaction splits the nuclear energy levels, with the splitting proportional to the product of the nuclear quadrupole moment (NQM) and the electric field gradient, according to
\begin{equation}
\label{Q-eq}
    \text{NQCC [MHz]} = 234.9647 \times Q(X)\,[\text{b}] \times q(X)\,[E_\mathrm{h}/a_0^2],
\end{equation}
where $X$ denotes the nucleus under investigation. Experimentally, this effect is quantified by the nuclear quadrupole coupling constant (NQCC), which directly reflects the interaction strength. Owing to the multiplicative nature of the quadrupole interaction, the determination of nuclear quadrupole moments relies on both accurate experimental measurements of nuclear quadrupole coupling constants and high-level calculations of electric field gradients (EFGs). The \textit{direct method} determines $Q$ for a single nucleus using Eq.~\eqref{Q-eq}, combining the measured NQCC with the calculated EFG. This approach, however, can be highly sensitive to small inaccuracies in the EFG, especially when its value is close to zero, as even minor errors in electron correlation contributions can lead to large relative deviations.\cite{HAIDUKE2026113055}  In contrast, the \textit{indirect method} estimates $Q$ by analyzing NQCCs across a series of molecules. Here, $Q$ is extracted from a linear regression of the experimental NQCCs against the corresponding calculated EFGs. By leveraging multiple molecules, this procedure reduces the impact of systematic errors and is generally more robust to small inaccuracies in the EFG of any individual system.\cite{Dufek1991,Belpassi2007} Nonetheless, obtaining accurate EFGs remains a significant challenge in quantum-chemical calculations. This is because the EFG is highly sensitive to several factors.
First, relativistic effects play a crucial role because the EFG depends on the electronic density close to the target nucleus, being proportional to $\braket{r^{-3}}$ (see, for instance, Refs. \citenum{TSAUEBOOK} and \citenum{Fabbro_JPCA_2025}) ; even small relativistic contractions or expansions of inner orbitals can significantly alter this expectation value. Second, electron correlation strongly influences the EFG.\cite{derevianko} According to Unsöld’s theorem,\cite{UNSOLD} closed atomic shells, such as we may expect from core orbitals, do not contribute to the EFG, unless they are polarized. Consequently, a proper description of electron correlation, particularly in the valence region, is essential to capture the correct anisotropy of the electronic density. Finally, the choice of basis set critically affects the results: since the EFG is determined by the electronic density near the nucleus and in the bonding region, an insufficiently flexible basis—especially in the core and polarization functions—can lead to large errors.\cite{KELLO1990641,Pyykkoe_TCA1997,Pernpointner2001_JCP,VANSTRALEN10072003,Haiduke2015,Shee2006,Fabbro2025}


The quadrupole moment of strontium (Z = 38) is of interest due to the widespread use of Sr isotopes in optical clocks, cold-atom experiments, and precision measurements probing fundamental symmetries.\cite{Barwood2004,2007PhDT.......201B,Bothwell2024} Among the naturally occurring isotopes, $^{84}$Sr, $^{86}$Sr, and $^{88}$Sr have nuclear spin zero and therefore $Q(\ce{^{84,86,88}Sr})=0$. The nuclear electric quadrupole moment is thus relevant only for the odd-$A$ isotopes and certain excited states. In particular, the ground states of $^{83}$Sr, $^{85}$Sr, and $^{87}$Sr, with spins $I=7/2^+$, $9/2^+$, and $9/2^+$, respectively, possess finite quadrupole moments.\cite{Stone2005} Sahoo \emph{et al.} combined the precisely measured hyperfine constant of the $4d\,^2D_{5/2}$ level in $^{87}$Sr$^+$ with a relativistic coupled-cluster calculation of $eq$ for that state, finding $Q(^{87}\mathrm{Sr}) = 0.305\pm 0.002\ \mathrm{b}$.\cite{Sahoo2006} This was a significant revision from earlier estimates by around 10\%,\cite{Martensson-Pendrill2002,Yu2004} and has been adopted as the recommended value by Pyykkö.\cite{Pyykko2008}
 More recently, Lu and co-workers revisited the NQM of $^{87}$Sr.\cite{Lu2019} Using high-precision multi-configuration relativistic Hartree-Fock calculations to obtain electric field gradients at the Sr position, combined with experimental hyperfine constants of the $5s5p\ ^3P_{1,2}$ states of neutral Sr, they obtained $Q(^{87}\mathrm{Sr}) = 0.328\pm 0.004$~b. This result differs from the previously recommended value but is in excellent agreement with earlier determinations by Mårtensson-Pendrill,\cite{Martensson-Pendrill2002} and by Yu \emph{et al.},\cite{Yu2004} and was proposed as the new reference value. Recently, an extraction of $Q(\ce{^87Sr})$ was reported by Tang,\cite{Tang2025} based on a relativistic hybrid configuration-interaction plus coupled-cluster (CI+CC) approach to calculate the electric field gradients for low-lying states of the neutral atom. By combining these EFGs with experimental electric quadrupole hyperfine-structure constants, Tang reported a value of $Q(\ce{^87Sr})=0.336\pm 0.004$ b, which is approximately 10\% higher than the currently recommended value.

The availability of nuclear quadrupole coupling constants for two molecules, $^{87}$Sr$^{16}$O ($-42.729(37)$~MHz)\cite{blom1992} and $^{87}$Sr$^{32}$S ($-21.959(85)$~MHz),\cite{ETCHISON200771} now enables the determination of the $^{87}$Sr nuclear electric quadrupole moment using the molecular approach. To our knowledge, no extraction of $Q(\ce{^87Sr})$ was performed from these values. Indeed, in 2017, Pyykkö emphasized that these molecular constants had not yet been utilized for determining $Q(\ce{^87Sr})$.\cite{Pyykkö2018}

The aim of this work is to extract the nuclear quadrupole moment of $^{87}$Sr from highly-accurate molecular relativistic relativistic calculations of the EFG, combined with the molecular data.~\cite{blom1992,ETCHISON200771}  


\section{Computational Details}

All computations were carried out using the DIRAC program package.\cite{DIRAC2020} The ID of the specific git commits of DIRAC utilized in this work are documented in the respective output files, which are made available in a dedicated online repository.\cite{Fabbro2026_Zenodo}
In this study, we mainly use the exact two-component molecular mean-field Hamiltonian (X2C$\mathrm{mmf}$).\cite{Sikkema2009}
The initial four-component Hartree--Fock (HF) calculations explicitly included scalar relativistic effects, spin--orbit interactions, as well as the Gaunt term and the SS integrals. Any calculations performed without one or more of these contributions are explicitly indicated in the text. For spin-free calculations, the spin-orbit interaction was eliminated as described in Ref.~\citenum{saue:spinfree}. For all molecular systems, we adopted the Gaussian nuclear charge distribution model, using the parameter set from Ref.~\citenum{Visscher:atomic}. 
For all atoms, we initially employed the dyall.ae4z basis set.\cite{Dyall2016} To assess the convergence of the EFG with respect to basis set extension, we subsequently augmented the Sr basis with two additional d-tight functions (exponents: 1.45205990$\times10^{4}$ and 6.52893664$\times10^{3}$) in an even-tempered manner, performing this augmentation at the SCF level (see Sec. \ref{Basis set and active space choice}).

We employed coupled-cluster (CC)\cite{CrawfordTDaniel2000AItC,BartlettRev20007} methods to compute the electric field gradient at the nucleus and to extract the nuclear quadrupole moment of $^{87}$Sr. 
For all CC schemes used in this work, the starting point was an iterative CCSD calculation performed with the RELCCSD module, which benefits from the exploitation of point-group symmetry.\cite{Visscher1995,DIRAC2020,Shee2006} Subsequently, triple excitations were accounted for using two complementary approaches. 
In the first, perturbative scheme, CCSD(T)\cite{Raghavachari1989,Bartlett1990,Stanton1997,Taube2008} and CCSD--T\cite{DEEGAN1994321} expectation values were evaluated using a finite-difference approach. The second scheme, that we denote CCSD$\tilde{\text{T}}$, is a composite method in which triple-excitation corrections are obtained analytically from fully iterative CCSDT calculations performed using smaller basis sets and active spaces. The resulting $T$ correction is then added to the CCSD value evaluated with a larger basis set and an active space sufficient to ensure convergence of the investigated property, thereby yielding the final CCSD$\tilde{\text{T}}$ result. This approach was employed in our previous work,\cite{Fabbro2025} where it enabled a reliable determination of the nuclear electric quadrupole moments of $^{27}$Al and $^{7}$Li.

Finally, we considered the vibrational corrections to the electronic values by employing the utility program VIBCAL, which is part of the DIRAC code\cite{DIRAC2020} and which was used previously for estimating the vibration contribution to the EFG.\cite{Fabbro2025} 

We determined $Q(\ce{^87Sr})$ by combining the NQCCs listed in Table~\ref{NQCCs} with the computed EFGs, according to Eq.~\eqref{Q-eq}.

\begin{table}[h!]
\centering
\begin{tabular}{lccc}
\toprule
Molecule & $\nu$ & NQCC (MHz) & Bond length (\AA) \\
\midrule
$^{87}$SrO & 0 & $-42.729$ \cite{blom1992} & 1.91983 \cite{NIST_Diatomic_Spectral_Database} \\
$^{87}$SrS & 0 & $-21.959$ \cite{ETCHISON200771} & 2.4405 \cite{NIST_Diatomic_Spectral_Database} \\
\bottomrule
\end{tabular}
\caption{Nuclear quadrupole coupling constants (NQCC) for SrO and SrS, vibrational quantum numbers $\nu$, and bond lengths.}
\label{NQCCs}
\end{table}

\subsection{CCSD(T) \& CCSD-T}
The CCSD(T) method,\cite{Raghavachari1989,Bartlett1990,Stanton1997,Taube2008} as implemented
in the RELCCSD module,\cite{Visscher1995,Visscher1996} provides a perturbative treatment of triple
excitations that incorporates the dominant contributions up to fifth order in many-body
perturbation theory. Owing to its favourable balance between accuracy and computational
cost, CCSD(T) has long been regarded as the \emph{de facto} standard for high-accuracy
quantum-chemical calculations and is frequently referred to as the ``gold standard’’ of
electronic-structure theory.\cite{CrawfordTDaniel2000AItC,BartlettRev20007} For many ground-state properties, this approach captures the
vast majority of the correlation effects associated with triple excitations, making it both
efficient and broadly applicable.

An alternative perturbative treatment is provided by the CCSD-T method, which extends the
CCSD(T) formalism by incorporating a wider class of disconnected triple excitations, again
up to fifth order in perturbation theory.\cite{DEEGAN1994321} From a formal standpoint,
CCSD-T is the most complete of the commonly used non-iterative triples corrections, since
it includes terms that are omitted in CCSD(T) and that become non-negligible in systems
exhibiting stronger electron correlation or near-degeneracy effects. Extensive benchmark studies have demonstrated the practical importance of these additional
terms for electric-field-gradient calculations.\cite{Visscher1998,DEEGAN1994321,HAIDUKE200795,
Haiduke2006,Haiduke2015,Dognon2025} 

Expectation values at the CCSD(T)/CCSD-T level were obtained using a finite-field
procedure, in which the property is evaluated as the numerical derivative of the electronic
energy with respect to an externally applied electric-field gradient. All finite-field
calculations were performed using relaxed Hartree–Fock orbitals in order to ensure a
consistent treatment of orbital relaxation effects across the different field strengths.

The reliability of the finite-field approach critically depends on the magnitude of the
applied perturbation. If the field is chosen too small, numerical noise arising from the
energy evaluation may obscure the physical response; if it is too large, higher-order
nonlinearities contaminate the numerical derivative. After performing several tests with different field strengths, we found that 
$\varepsilon = \pm 10^{-7}$ offers an optimal balance between numerical stability and 
maintaining the linear-response regime. Such values were successfully used in other NQM determinations.\cite{Haiduke2006,HAIDUKE200795,Santiago2014,Gusmao2019} The correlation contribution to the
HF reference EFG is obtained from the antisymmetric energy difference,
\begin{equation}
    q_\mathrm{el, corr}
   \approx
   \frac{E_\mathrm{corr}(\varepsilon) - E_\mathrm{corr}(-\varepsilon)}{2\varepsilon}
   \Bigg|_{\varepsilon = 0},
\end{equation}
which isolates the first-order response of the correlation energy with respect to the external field. The final EFG value is obtained by adding the above correlated term, $q_\mathrm{el, corr}(X)$, to the HF value $q_{\text{el,HF}}(X)$, since it has been observed that the electron correlation contribution to the EFG is essentially linear.\cite{Pernpointner1998_A,Pernpointner1998CsF}

The definition of an appropriate active space is a crucial prerequisite for obtaining
reliable coupled-cluster results on heavy diatomic systems. In the present work, the active
space was selected by in-depth analysis of the convergence behavior of the EFG in the SrO
molecule. To this end, we performed a systematic series of CCSD calculations in which the number
of virtual orbitals included in the correlation treatment was gradually increased while
keeping all electrons correlated.  After an analysis (see below, Figure \ref{fig:efg-vs-spinor-energy} ), an energy cutoff of 50~$E_{h}$ was adopted as a balanced and computationally efficient choice for defining the active virtual space in all subsequent
calculations for both SrO and SrS. 

CCSD, CCSD(T) and CCSD-T calculations were performed on the CALMIP supercomputing center (Centre de Calcul Intensif des Midi‑Pyrénées) in Toulouse, France. The calculations were run on the Olympe partition of CALMIP, which comprises high‑performance compute nodes each equipped with dual Intel Skylake 6140 processors at 2.3 GHz (36 physical cores per node) and 192 GB of DDR4 main memory.

\subsection{CCSD$\tilde{\text{T}}$}
The CCSD$\tilde{\text{T}}$ composite method was employed in our previous work.\cite{Fabbro2025} In summary, it is based on evaluating the T corrections using a smaller basis set (dyall.v3z basis set\cite{Dyall2016} in our case) and a reduced active space,
\begin{equation}
    T_{\text{iter}} = \braket{q}_{\text{CCSDT}}^{\text{dyall.v3z}} - \braket{q}_{\text{CCSD}}^{\text{dyall.v3z}},
\end{equation}
This correction is then added to the reference value obtained at the CCSD level 
\begin{equation}
    \braket{q}_{\text{CCSD$\tilde{\text{T}}$}} =\braket{q}_{\text{CCSD}}^{\text{opt-basis}} + T_{\text{iter}},
\end{equation}
with a basis guaranteeing the convergence of both the EFG, as well as the active space.
The fully iterative CCSDT method\cite{Lee1984,Urban1985,Noga1987,Brandejs2025} and the respective expectation values counterpart,\cite{Gauss_JCP2002,Fabbro2025} is currently available in the ExaCorr \cite{Pototschnig2021} module of DIRAC, thanks to the \underline{tenpi} code generator,\cite{Brandejs2025} based on a scheme introduced by Kallay and Surjan.\cite{Kallay2001} The CC implementation is unrestricted and can, in principle, be applied to any system well described by a single-reference wave function; in this work, however, all calculations are based on Kramers-restricted HF orbitals.\cite{Saue_JCP1999}    

 For the CCSDT calculations, correlation was restricted to the four highest-lying occupied Kramers pairs (i.e two spinors related by time-reversal symmetry); the same choice was adopted for both SrO and SrS. In addition, 97 virtual Kramers pairs were included in the correlation treatment for both molecules. Using the dyall.v3z basis set, this choice corresponds to an energetic cutoff of approximately $3.8\,E_\mathrm{h}$ for the virtual space in SrO, while a slightly higher cutoff of about $4.3\,E_\mathrm{h}$ was required for SrS. This choice was motivated both by computational cost considerations and by the need to avoid truncating virtual shells midway, which could otherwise lead to rapid and unphysical oscillations in the computed EFG values.\cite{Fabbro2025}

CCSDT calculations were performed on the Genoa partition of the French Adastra supercomputer, which packs 192 cores of a dual AMD EPYC 9654 chip with 768 GB of random access memory (RAM) into each node. The 544 nodes of Genoa partition are connected with 4x25GB/s Cray Slingshot-11 interconnects.

\subsection{Massively parallel tensor operations}

The T corrections are obtained from the fully-iterative CCSDT method,\cite{Lee1984,Urban1985,Noga1987} which was made possible thanks to the recent adoption of a standard tensor contraction interface into DIRAC. The Tensor Algebra Processing Primitives (\underline{TAPP}) is a newly published standard\cite{brandejs2026tensoralgebraprocessingprimitives} C-API (application programming interface) enabling programs to call one of the multiple supported tensor libraries without the need of changing the code when switching between the libraries. DIRAC is the first quantum chemistry package to use TAPP. Specifically, TAPP opened the way for the support of cutting-edge tensor libraries like TBLIS\cite{Matthews2018} and cuTENSOR,\cite{cuTENSOR} built on the shared-memory parallelization model. One of the present authors (J.B.) have further extended TAPP to a distributed memory multi-node variant.\cite{tappctf} This allows any program using TAPP to operate on tensors too large to fit into the RAM of one computational node. Such tensor operations require specialized distributed memory libraries like the Cyclops Tensor Framework (CTF),\cite{Solomonik2014} used in the present work. CTF is built around an advanced communication-minimizing parallelization scheme. Recently, we have used CTF to demonstrate excellent weak scaling performance with CCSDT and CCSDTQ tensor contractions on up to 130 Frontier supercomputer nodes.\cite{Brandejs2025}

The present work is the first to implement CTF-based massive parallelization with the new molecular properties module in DIRAC,\cite{Fabbro2025} and the first to use the French national HPC platform Adastra for such purpose.


TAPP is a C-API which passes calls from DIRAC Fortran to Cyclops C++ code. To enable distributed memory operations, the TAPP specification had to be modified to eschew direct pointers to tensor data, as the data are scattered across multiple nodes. Instead, we extended the tensor metadata object "tensor\_info" to include a universally unique identifier (UUID) of the tensor, which is used to retrieve the corresponding CTF tensor instantiated in our C++ worker code that wraps CTF. Except for the small modification, the wrapper implements the standard TAPP interface.

Additional development was required to align the different parallelization schemes. In DIRAC, the master MPI process assigns tasks to worker processes, whereas CTF assumes a symmetric parallelization with the same call from all the processes simultaneously. Our wrapper of CTF implements message passing from the master process of DIRAC to dedicated CTF worker processes. 

All above tensor interface components are not restricted to DIRAC. The code is portable and openly accessible under a permissive license to anyone in the community seeking to use massively parallel tensor libraries in their application code through a fixed API.\cite{tappctf}



\section{Results and Discussion}
\subsection{Bonding analysis}
Both molecules in this study, SrO and SrS, display ionic bonding, as inferred from projection analysis.\cite{Dubillard_JCP2006} The EFG at the Sr nucleus is thereby not induced by covalent bonding, but by polarization by the partner anion. Using Intrinsic Atomic Orbitals,\cite{Knizia_JCTC2013} which assures that the HF molecular orbitals (MOs) are fully spanned by the precomputed occupied orbitals of the constituent atoms, hence removing any polarization contribution, we find that the charge of Sr is +1.98e and +1.81e in SrO and SrS, respectively. Interestingly, in SrO we observe a mixing of Sr 4p and O 2s orbitals in canonical MOs, but upon localization according to the Pipek--Mezey criterion,\cite{Pipek-Mezey_JCP1989} it goes away. In the terminology of Neidig \textit{et al.} this is overlap-driven covalency,\cite{Neidig_CCR2013} but clearly does not constitute true chemical bonding.

\subsection{Basis set and active space choice}
\label{Basis set and active space choice}
Basis sets are optimized for energies rather than properties; it is therefore necessary to verify the stability of the molecular property with respect to basis-set extension. The convergence test was performed only on the SrO molecule, which is the most ionic system. As discussed in our previous work,\cite{Fabbro_JPCA_2025} the EFG at the X nucleus is predominantly determined by the orbitals centered on that nucleus. Furthermore, we showed that contributions from surrounding nuclei largely cancel out, leaving orbital polarization by the ligands as the dominant effect. For this reason, our study focuses on the most polarized systems. The basis-set study was carried out at the HF-level, since we assume that the basis sets are already well equipped with correlation functions.

In our augmentation study, we progressively added higher–angular–momentum functions (\textit{p}, \textit{d}, \textit{f}, etc.) to the uncontracted \texttt{dyall.ae4z} basis set \cite{Dyall2016} until the change in the computed electric–field gradient became negligible. As shown in Table~\ref{tab:augmentation_summary}, the most substantial corrections were observed upon the addition of \textit{d}-type functions, particularly up to 2\textit{d}. Notably, the addition of the third set of \textit{d}-type functions (3\textit{d}) leads to a relative change of just 0.012\%, which is well below both the experimental uncertainty \(\delta_{\mathrm{exp}} = 0.087\%\). Therefore, we chose to truncate the augmentation at 3\textit{d}, as any further addition of \textit{d}-functions did not result in consistent improvements and would only increase computational cost without gaining accuracy. The subsequent addition of f-tight functions did not modify the EFG.

\begin{table}[h!]
\centering
\footnotesize
\setlength{\tabcolsep}{6pt}
\renewcommand{\arraystretch}{0.9}
\begin{tabular}{c|c|c|c}
EFG ref.\ ($E_h/e a_0^2$) & Exp. & EFG ($E_h/e a_0^2$) & Change (\%) \\
\hline
-0.50244 & --  & --             & --        \\
              & 1p  & -0.50245  & 0.002     \\
\hline
-0.50245 & --  & --             & --        \\
              & 1d  & -0.50040  & 0.405     \\
              & 2d  & -0.49736  & 0.608     \\
              & 3d  & -0.49742  & 0.012     \\
\hline
 -0.49736 & 1f  & -0.49740  & 0.007     \\
              & 2f  & -0.49738  & 0.005     \\
\end{tabular}
\caption{Effect of p-, d-, and f-type basis augmentations on the EFG at the Sr position in SrO at the DCG-HF level.}
\label{tab:augmentation_summary}
\end{table}

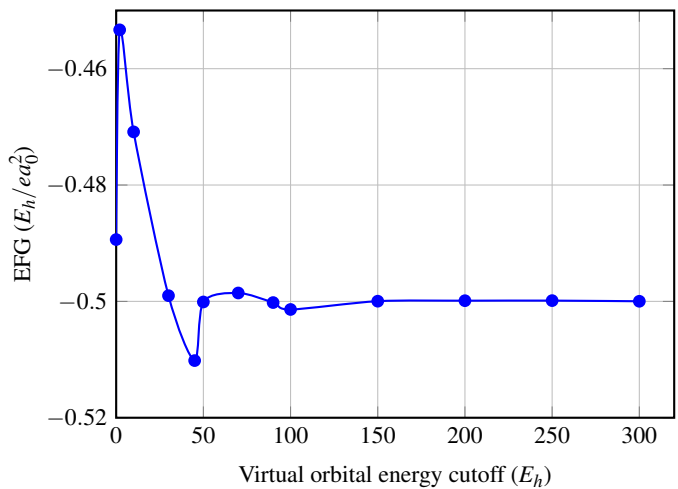
\begin{figure}[h!]
  \centering
  \begin{tikzpicture}
    \begin{axis}[
      width=9cm,
      height=7cm,
      xlabel={Virtual orbital energy cutoff (\(E_h\))},
      ylabel={EFG (\(E_h/e a_0^2\))},
      grid=major,
      thick,
      tick label style={font=\small},
      label style={font=\small},
      legend style={font=\small, at={(0.98,0.98)}, anchor=north east},
      xmin=0, xmax=320,
      ymin=-0.52, ymax=-0.45
      ]
      \addplot[
        blue,
        mark=*,
        smooth,
        thick
      ] coordinates {
        (0, -0.4893744442)
        (2, -0.4533255873)
        (10, -0.4708633420)
        (30, -0.4990170175)
        (45, -0.5101771133)
        (50, -0.5001011376)
        (70, -0.4985571449)
        (90, -0.5002027344)
        (100, -0.501391055)
        (150, -0.4999670333)
        (200, -0.4998826060)
        (250, -0.4998688132)
        (300, -0.4999809849)
      };
    \end{axis}
  \end{tikzpicture}
  \caption{Electric field gradient (EFG) of the SrO molecule as a function of the virtual orbital energy cutoff, computed at the equilibrium geometry at the DCG-CCSD level.}
  \label{fig:efg-vs-spinor-energy}
\end{figure}

For the correlation space, we adopted an energy-based virtual orbital cutoff rather than orbital-specific selection to avoid abrupt electric field gradient oscillations, as discussed in our previous work.\cite{Fabbro2025} We considered the SrO molecule and increased systematically the number of virtual orbitals included in the active space by energy cutoff. Looking at the plot reported in Figure \ref{fig:efg-vs-spinor-energy}, we observed that the EFG value converges as the number of virtual spinors increases. In particular, once the energy threshold reaches 150~$E_h$, further inclusion of virtual spinors has a negligible effect: increasing the cutoff from 150~$E_h$ to 200~$E_h$ changes the EFG by only 0.02\%, which is smaller than the experimental uncertainty. However, still the energy threshold 150 $E_h$ is computational demanding. By applying an energy cutoff of 150~$E_h$ for the virtual orbitals, we obtain a value of $eq = -0.49997$~$E_h/ea_0^2$. In comparison, a more restrictive cutoff of 50~$E_h$ yields $eq = -0.50010$~$E_h/ea_0^2$. The relative difference between these two values is approximately 0.03\%, which is notably smaller than the experimental relative uncertainty reported for the nuclear quadrupole coupling constants of SrO and SrS. This indicates that, for both molecular systems, adopting a cutoff of 50~$E_h$ for the virtual orbitals is sufficient to guarantee that the truncation error in the EFG remains negligible and does not affect the subsequent determination of the nuclear quadrupole moments.

\subsection{Electric field gradient analysis}
The data in Table~\ref{tab:efg_contributions} reveal clear and quantifiable trends in how relativistic and correlation effects affect the electric field gradients and nuclear quadrupole moments of SrO and SrS.  
Starting from the non-relativistic HF reference, the inclusion of relativistic effects (4c-DC-HF) reduces the magnitude of the EFG by 2.0\% in SrO and by 14.0\% in SrS, resulting in increases of 2.0\% and 16.3\% in the corresponding quadrupole moments.  Introducing the Gaunt interaction (4c-DCG-HF) produces an additional reduction of 1.6\% (SrO) and 1.4\% (SrS) in the EFG magnitudes, with the associated $Q$ values increasing by the same percentages. This trend is consistent with previous observations on heavy-element systems. Pernpointner~\cite{MPernpointner_2002} reported that for light systems such as AlF and AlCl the Gaunt contribution is negligible, whereas in strongly relativistic systems like TlH, it leads to a pronounced EFG shift of approximately $0.1~E_h/ea_0^2$. Compared to these studies, SrO and SrS represent intermediate cases: the Gaunt term has a small but non-negligible effect, larger than in light systems but smaller than in strongly relativistic systems, reflecting the moderate relativistic character of strontium. One should note that the gauge-dependent term in the full Breit operator often contributes negligibly;\cite{Pernpointner2004} since the Gaunt contribution is only of the order of $1.5$--$1.6\%$ for both systems, we expect this additional term to be even smaller. \cite{Belpassi2007}

Electron correlation at the CCSD level modifies the relativistic HF picture in a system-dependent way. For SrO, CCSD increases the magnitude of the EFG by 2.2\% relative to 4c-DCG-HF, whereas SrS shows a substantially larger 14.2\% decrease in magnitude.  
The quadrupole moments reflect these changes, decreasing by 2.1\% in SrO and increasing by 16.7\% in SrS. This contrasting behaviour highlights the different sensitivity of the two chalcogenides to dynamical correlation.

Triple excitations have a significant quantitative impact when assessed relative to CCSD.  
The perturbative triples correction, CCSD(T), reduces the EFG magnitude by 9.8\% in SrO and by 2.4\% in SrS compared to CCSD. The associated quadrupole moments decrease by 8.9\% (SrO) and 2.3\% (SrS), confirming that perturbative triples systematically lower both EFG and $Q$.

The CCSD-T method partially counteracts this trend.  
Relative to CCSD, the EFG magnitudes change by only 6.4\% in SrO and 2.7\% in SrS, corresponding to quadrupole variations of $+2.3\%$ (SrO) and $+2.8\%$ (SrS).  
These results indicate that CCSD-T introduces corrections of opposite sign to CCSD(T).

The CCSD$\tilde{\text{T}}$ method exhibits the largest deviations from the CCSD reference among the correlated approaches considered. 
For SrO, the inclusion of iterative triples leads to an increase in the magnitude of the EFG by approximately 4.3\% (i.e., a more negative value) relative to CCSD, while the nuclear quadrupole moment is reduced by about 4.2\%. In contrast, for SrS the effect of iterative triples is more pronounced in the opposite direction: the EFG magnitude is reduced by roughly 9.3\%, whereas the quadrupole moment increases by about 10.3\% with respect to CCSD.

Overall, the results demonstrate that relativistic effects modify the EFGs by up to about 15\%, while dynamical electron correlation may further alter them by up to 17\%. 
The inclusion of triple excitations, in particular through the CCSD$\tilde{\text{T}}$ treatment, affects both EFGs and quadrupole moments by up to $\sim$10\%--13\%, with a notably stronger impact observed for SrO than for SrS. 
These findings confirm that reliable predictions of quadrupole moments in alkaline-earth chalcogenides require a fully relativistic framework combined with a high-level correlated treatment.
\begin{table}[H]
\centering
\begin{tabular}{l
                S[table-format=-1.5] 
                S[table-format=1.5] 
                S[table-format=-1.5] 
                S[table-format=1.5]}
\hline\hline
& \multicolumn{2}{c}{SrO} & \multicolumn{2}{c}{SrS} \\
Method & {EFG} & {Q} & {EFG} & {Q} \\
\hline
HF (no-rel)               & -0.50717 & 0.35856 & -0.37480 & 0.24935 \\
4c-DC-HF                  & -0.49730 & 0.36568 & -0.32225 & 0.29001 \\
4c-DCG-HF                 & -0.48938 & 0.37160 & -0.31780 & 0.29407 \\
X2Cmmf-G-CCSD$^{a)}$      & -0.50010 & 0.36363 & -0.27254 & 0.34291 \\
X2Cmmf-G-CCSD(T)$^{a)}$          & -0.54926 & 0.33109 & -0.27905 & 0.33491 \\
X2Cmmf-G-CCSD-T$^{a)}$           & -0.53210 & 0.34176 & -0.26518 & 0.35242 \\
X2Cmmf-G-CCSD$\tilde{\text{T}}$$^{b)}$ & -0.52182 & 0.34850 & -0.24718 & 0.37809 \\
\hline\hline
\end{tabular}
\caption{EFG (in $E_{\mathrm{h}}/e a_{0}^{2}$) and quadrupole moments (in b) for SrO and SrS at the equilibrium geometries.
${a)}$ calculations were performed with the \textit{opt} basis set, correlating all occupied orbitals and all virtual orbitals up to 50~$E_{\mathrm{h}}$.
${b)}$ The iterative T contribution was obtained using the \texttt{dyall.v3z}, correlating the four highest-lying occupied Kramers pairs and the lowest 97 virtual Kramers pairs.}
\label{tab:efg_contributions}
\end{table}

\subsection{Quadrupole moment analysis}
In Table~\ref{tab:efg_mad} we list the quadrupole moments $Q$ of \ce{^{87}Sr} in SrO and SrS obtained at different levels of theory. In addition, we list the average value $\overline{Q}$, and the corresponding Mean Absolute Deviation (MAD).\cite{mad} Ideally, the MAD should be zero since the quadrupole moment is a nuclear property independent of the chemical environment.
The Table shows a clear hierarchy in the ability of different levels of theory to produce internally coherent quadrupole moments for the two chalcogenides.  
The non-relativistic HF values exhibit the largest deviation (MAD = 0.0546 b), reflecting the well-known imbalance in describing heavy-element systems without relativistic and correlation effects. Introducing relativistic effects (4c-DC-HF) reduces the MAD by approximately 30\%, while the inclusion of the Gaunt term yields a very similar level of internal consistency. 

Electron correlation dramatically enhances consistency. At the CCSD level, the MAD drops to 0.0102 b, representing an 80\% reduction relative to the relativistic HF values. This demonstrates that dynamical correlation is essential to obtain comparable quadrupole moments for SrO and SrS.  
Among all methods considered, CCSD(T) yields the smallest deviation (MAD = 0.00191 b), almost an order of magnitude smaller than CCSD and more than 25 times smaller than non-relativistic HF. This confirms that perturbative triples not only improve absolute accuracy (as shown in the EFG analysis) but also produce highly balanced predictions across different chemical environments.

The CCSD-T method gives a slightly larger MAD (0.00567 b), still significantly smaller than at the HF or even CCSD level, but clearly less uniform than CCSD(T).  Finally, the CCSD$\tilde{\text{T}}$ method yields a MAD of 0.01480~b, indicating that, while this approach introduces substantial and physically meaningful corrections to the absolute EFGs and quadrupole moments, it also amplifies the difference between SrO and SrS. This results in a slightly less homogeneous description across the two systems. This behavior may be traced back to the limited virtual space used in these calculations. As we have observed in Figure \ref{fig:efg-vs-spinor-energy}, converged EFGs require a virtual space cutoff of at least approximately 50~$E_\mathrm{h}$, which is not achievable in the current CCSDT calculations due to computational constraints. Consequently, the contribution of the triple excitations cannot be fully captured by the restricted virtual space employed here, leading to a partial representation of their effect. 

The MAD analysis shows that relativistic HF reduces the inconsistency between the two systems, electron correlation is crucial to obtain tightly clustered quadrupole moments, and perturbative triples [CCSD(T)] provide the most balanced description across the SrX series.

\begin{table}[H]
\centering
\label{tab:efg_mad}
\begin{tabular}{l
                S[table-format=1.5]
                S[table-format=1.5]
                S[table-format=1.5]
                S[table-format=1.5]}
\toprule
Method & {$Q(\ce{^87Sr})[\ce{SrO}]$} & {$Q(\ce{^87Sr})[\ce{SrS}]$} & {$\overline{Q}(\ce{^87Sr})$} & {MAD} \\
\midrule
\hline
HF (no-rel)               & 0.35856 & 0.24935 & 0.30396 & 0.05461 \\
4c-DC-HF                  & 0.36568 & 0.29001 & 0.32785 & 0.03784 \\
4c-DCG-HF                 & 0.37160 & 0.29407 & 0.33284 & 0.03876 \\
X2Cmmf-G-CCSD             & 0.36363 & 0.34291 & 0.35327 & 0.01018 \\
X2Cmmf-G-CCSD(T)          & 0.33109 & 0.33491 & 0.33300 & 0.00191 \\
X2Cmmf-G-CCSD-T           & 0.34176 & 0.35242 & 0.34709 & 0.00567 \\
X2Cmmf-G-CCSD$\tilde{\text{T}}$ & 0.34850 & 0.37809 & 0.36330 & 0.01480 \\
\hline
\bottomrule
\end{tabular}
\caption{Quadrupole moments $Q$ of \ce{^{87}Sr} in SrO and SrS, their average $\overline{Q}$, and the corresponding Mean Absolute Deviation (MAD) across different relativistic and correlated methods. All values are reported in barns.}
\end{table}

From the above analysis, we take the CCSD(T) value as the most reliable estimate. This conclusion is supported by its exceptionally low MAD, which indicates the most internally consistent description across the SrX series, together with the well-established accuracy of the perturbative triples correction in capturing the dominant contributions beyond CCSD. Unlike the CCSD$\tilde{\text{T}}$ method, here we were able to employ a larger active space, including all occupied orbitals and all virtual orbitals up to 50~$E_\mathrm{h}$.

Finally, vibrational corrections were included and computed at the CCSD level using the utility program \texttt{VIBCAL},\cite{DIRAC2020} accounting for cubic anharmonicities. The results are reported in Table \ref{vib-results}.

The vibrational analysis was performed by constructing, for each molecule, both a potential energy surface (PES) and a corresponding property surface (PS). These surfaces were generated by sampling multiple points along the molecular coordinate at the CCSD level, employing the optimized basis set and the converged active space defined by an energy cutoff of 50~$E_h$ for the virtual orbitals, as discussed in Sec.\ref{Basis set and active space choice}. In addition to the equilibrium geometry, four displaced geometries were considered, obtained by varying the molecular coordinate by $\pm 0.005$~\text{\AA}. 
This resulted in a total of five geometries per molecule used to construct the PES and PS required for the vibrational treatment. 
One may question the use of the CCSD method for estimating vibrational contributions, given that the electronic EFG values were obtained at the CCSD(T) level. It should be noted, however, that for each molecule this analysis would require at least five CCSD(T) property calculations per geometry. When further accounting for the two finite-field points needed, two molecular systems, and five geometries per molecule, the total number of CCSD(T) calculations would amount to 20, rendering such an approach computationally too expensive. Importantly, the vibrational contributions reported in Table~\ref{vib-results} are relatively small, amounting to 0.00905~$E_h/e a_0^2$ for SrO and 0.00296~$E_h/e a_0^2$ for SrS. These values are of the same order of magnitude as the estimated uncertainties in the electronic EFGs for both molecules (see Section~\ref{Error estimation}). Consequently, any potential improvement gained by performing the vibrational analysis at the CCSD(T) level instead of CCSD is expected to be minor and well within the overall uncertainty of the final EFG values.
\begin{table}[H]
\centering
\begin{tabular}{lccccc}
\toprule
\hline
Molecule & $\nu$ & EFG$_\text{CCSD(T)}$ & $\Delta$EFG$_\text{vib}$ & EFG$_\text{final}$ & $Q_\text{final}(\ce{^87Sr})$ \\
\midrule
SrO & 0 & -0.54926 & 0.00905& -0.54021& 0.33663 \\
SrS & 0 & -0.27905 & 0.00296& -0.27757& 0.33670\\
\hline
\bottomrule
\end{tabular}
\caption{CCSD(T) EFGs, vibrational corrections, final EFGs and final quadrupole moments for SrO and SrS. All the values are reported in $E_h/ea_0^2$.}
\label{vib-results}
\end{table}

\subsection{Error estimation}
\label{Error estimation}
In order to determine the uncertainty of our results, we identified the sources of error in the calculation of the EFG for the two molecules SrO and SrS. 
\begin{enumerate}
    \item \textbf{Basis set}: From our augmentation study (Table~\ref{tab:augmentation_summary}), the largest contributions to the EFG at the Sr position arise from the addition of \textit{d}-type functions, particularly the first two sets (1d: 0.405\%, 2d: 0.608\%). After the third set of \textit{d}-functions (3d), the EFG changes by only 0.012\%, and the subsequent inclusion of higher \textit{d} and \textit{f}-type functions results in variations below 0.02\%, which are likely dominated by numerical noise rather than systematic basis set incompleteness. From these observations, we estimate the error due to the incomplete basis set to be approximately $\Delta_{\mathrm{BSIE}} \sim 6 \times 10^{-5}\ E_h/ea_0^2 \quad (\sim 0.012\%)$

\item \textbf{Active space}: At the CCSD level we correlated all electrons and included virtual orbitals up to 50~$E_h$. Based on this convergence behavior, we estimate the error introduced by truncating the active space at 50~$E_h$ to be $\Delta_{\mathrm{AS}} \sim 1.3 \times 10^{-4}\ E_h/ea_0^2 \quad (\sim 0.03\%)$.
\item \textbf{Gauge term effects in the full Breit two-electron interaction}: In our calculations, we include the Gaunt term, which represents the magnetic interactions between electrons at the leading order in the Breit operator. 
Belpassi and co-workers performed a detailed analysis of the full, untruncated Breit interaction, and observed that the gauge terms reduce the magnetic (Gaunt) contribution by roughly one quarter in the AuF molecule.~\cite{Belpassi2007} 
Since Sr is considerably lighter than Au, we expect relativistic terms to be less pronounced. Therefore, applying the AuF ratio to SrO likely provides a conservative upper bound for the uncertainty.
The resulting estimate of the gauge contribution is 0.002 $E_{\mathrm{h}}/ea_0^2$ for SrO and 0.001 $E_{\mathrm{h}}/ea_0^2$ for SrS, respectively.
\item \textbf{High-order correlation effects}: Assessing the impact of electron correlation beyond the CCSD(T) level is inherently difficult, as the contribution of higher-order excitations is not straightforward to quantify. While CCSD(T) typically offers a marked improvement over CCSD, comprehensive benchmarks at levels exceeding CCSD(T) are relatively rare. Following the reasoning of Van Stralen and Visscher,\cite{VANSTRALEN10072003} who build upon earlier work by Bieroń \textit{et al.},\cite{Bieron2001} we adopt a conservative estimate that higher-order excitations contribute approximately 1\% to the EFG. This value is therefore used as an uncertainty associated with neglected correlation effects beyond CCSD(T).
\item \textbf{Higher-order anharmonic contributions:} Contributions beyond cubic order are expected to be at least one order of magnitude smaller than the cubic anharmonic contributions, which in our case were on the order of $10^{-4}~E_h/ea_0^2$, and are therefore anticipated to be around $10^{-5}~E_h/ea_0^2$.
\end{enumerate}

Based on the above analysis, we report the CCSD(T) values with their estimated uncertainties as $-0.54021 \pm 0.00585\ E_{\mathrm{h}}/ea_0^2$ for SrO and $-0.27757 \pm 0.00302\ E_{\mathrm{h}}/ea_0^2$ for SrS. Since SrO and SrS provide two independent estimates of the same nuclear quadrupole moment, the final value of $Q(\ce{^87Sr})$ was obtained as an inverse-variance–weighted average of the two molecular determinations, giving $Q(\ce{^87Sr})=0.33666\pm 0.00258$ b. This result is slightly above the value reported by Lu and co-workers,\cite{Lu2019} but it is in excellent agreement with the recent determination by Tang,\cite{Tang2025} $Q(\ce{^87Sr})=0.336\pm 0.004$ b, as both values are fully consistent within their respective uncertainties.


\section{Conclusions}

In this work, the nuclear electric quadrupole moment (\(Q\)) of \({}^{87}\mathrm{Sr}\) was determined using a highly accurate molecular relativistic approach that combines experimentally measured nuclear quadrupole coupling constants (NQCCs) of SrO and SrS with electric field gradients (EFGs) computed at various coupled-cluster levels, including CCSDT, CCSD-T, and CCSD$\tilde{\text{T}}$ within an exact two-component molecular mean-field Hamiltonian (X2Cmmf). The treatment explicitly incorporates important effects such as relativistic effects, the Gaunt two-electron interaction and vibrational corrections, while carefully optimizing basis sets and active spaces to ensure convergence and reliability. Electron correlation is captured comprehensively through the inclusion of triple excitations, assessed via perturbative and iterative schemes, thanks to the recent multi-node extension of our module for high-order CC energies and expectation values, made possible by the TAPP interface connecting the Cyclops CTF library with the DIRAC program. We have shown that the inclusion of triple excitations (either perturbative or iterative) is essential for the accurate determination of the EFG in SrO and SrS.

Since we used three different methods to estimate the contribution of triple excitations (CCSD(T), CCSD-T, and CCSD$\tilde{\text{T}}$), we employed the MAD to determine which provided the best value. We found that the best result is produced by CCSD(T). This greater internal consistency of CCSD(T) compared to CCSD$\tilde{\text{T}}$ is likely due to the fact that the active space we employed is sufficiently large to ensure convergence of the EFG with respect to further enlargement. In contrast, with CCSD$\tilde{\text{T}}$, although the multi-node extension allowed for an increase in the active space compared to the single-node version, it is still limited both in the choice of electrons to correlate and in the virtual space. This limitation is also certainly due to the fact that point-group symmetry is not currently implemented in the ExaCorr module of DIRAC, where we perform high-order CC calculations. In contrast, CCSD(T) is implemented in the RELCCSD module of DIRAC, which benefits from point-group symmetry. A future development we are actively pursuing is the introduction of point-group symmetry in ExaCorr.

Our best CCSD(T) value is $Q(\ce{^{87}Sr}) = 0.33666 \pm 0.00258$~b, slightly above the value proposed by Lu and co-workers,\cite{Lu2019} $Q(\ce{^{87}Sr}) = 0.328 \pm 0.004$~b. Lu et al. employed a multiconfiguration Dirac-Hartree-Fock (MCDHF) and relativistic configuration interaction (RCI) framework on the neutral Sr atom, focusing on the $5s5p\,{}^3P_{1,2}$ states and systematically accounting for electron correlations via single and double excitations from an extensive multireference space. However, they did not explicitly include triple excitations, which are essential for accurate determination of EFGs, as we have shown.

In contrast, Tang employed a hybrid configuration-interaction plus coupled-cluster (CI+CC) approach on the low-lying states of the neutral Sr atom to compute electric-field gradients, which were then combined with experimental hyperfine constants to extract $Q(\ce{^{87}Sr}) = 0.336\pm 0.004$~b.\cite{Tang2025} This method explicitly accounts for core-core, core-valence, and valence-valence correlations and incorporates relativistic effects. While Tang's approach is atom-centered and relies on atomic CI+CC calculations, our method instead uses molecular EFGs from two different molecules, SrO and SrS, combined with experimentally measured NQCCs, providing two independent estimates of $Q(\ce{^{87}Sr})$ in distinct chemical environments. By probing the nucleus within a chemical context, our approach naturally incorporates bonding effects and electron correlation in a realistic molecular setting, allowing us to cross-check the consistency of $Q(\ce{^{87}Sr})$ across multiple environments and thereby reduce potential systematic errors that may arise in purely atomic calculations. Despite these methodological differences, both atomic and molecular approaches converge to essentially the same value within uncertainties, highlighting the robustness and reliability of current high-accuracy theoretical determinations of the nuclear quadrupole moment of \ce{^{87}Sr} and providing a strong cross-validation between atomic and molecular data.



\section{Acknowledgments}

This project was funded by the European Research Council (ERC) under the European Union’s Horizon 2020 research and innovation program (grant agreement ID:101019907). This work was performed using HPC resources from CALMIP (Calcul en Midi-Pyrenées; Grant 2024-P13154). 
This work was supported by a French government grant managed by the Agence Nationale de la Recherche under the "Investissements d'avenir" program (reference "ANR-21-ESRE-0051"). This work was granted access to the MesoNET resources center and the MesoNET Project under the allocation M24070. 
This work was granted access to the HPC resources of CINES under the allocation 2025-A0190801859 made by GENCI. 
This research used resources of the Oak Ridge Leadership Computing Facility 
at the Oak Ridge National Laboratory, which is supported by the Office of Science 
of the U.S. Department of Energy under Contract No. DE-AC05-00OR22725. 

We dedicate this paper to the fond memory of John Stanton.

\section*{Author declarations}
\subsection*{Conflict of Interest}
The authors declare that they have no conflict of interest. 
\section*{Data Availability Statement}
The data that support the findings of this study will be made openly available in ZENODO.\cite{Fabbro2026_Zenodo} The corresponding computational outputs will be uploaded once the paper has been accepted.
\clearpage
\onecolumngrid
\section*{TOC}
\begin{center}
\begin{tikzpicture}[
    font=\sffamily,
    >=Stealth,
    atom/.style={
        circle, 
        shading=ball, 
        minimum size=1.1cm, 
        text=white, 
        font=\bfseries\sffamily, 
        drop shadow={opacity=0.3, shadow xshift=1pt, shadow yshift=-1pt}
    },
    bond_base/.style={line width=0.25cm, draw=gray!50, cap=round},
    bond_high/.style={line width=0.08cm, draw=gray!20, cap=round, shorten <=0.05cm, shorten >=0.05cm, yshift=0.06cm},
    methodbox/.style={
        rectangle, 
        rounded corners=4pt, 
        fill=tech_dark, 
        text=white,
        blur shadow={shadow blur steps=10, shadow opacity=20},
        align=center,
        inner sep=10pt,
        font=\small
    },
    resultbox/.style={
        rectangle, 
        rounded corners=4pt, 
        fill=white, 
        draw=gray!30, 
        very thick,
        blur shadow={shadow blur steps=10},
        align=center,
        inner sep=12pt
    },
    flowline/.style={
        ->, 
        line width=2pt, 
        draw=accent_blue!80, 
        shorten >=3pt,
        postaction={decorate},
        decoration={markings, mark=at position 1 with {\arrow{Stealth[length=3mm]}}}
    }
]

    
    \coordinate (SrO_center) at (0.5, 2.5);
    
    \draw[bond_base] (0, 2.5) -- (1.8, 2.5);
    \draw[bond_high] (0, 2.5) -- (1.8, 2.5);
    
    \node[atom, ball color=atom_sr] (Sr1) at (0, 2.5) {Sr};
    \node[atom, ball color=atom_o] (O1) at (1.6, 2.5) {O};
    
    \node[above=0.3cm of O1, text=accent_blue!80!black, font=\bfseries\footnotesize] {SrO};
    \node[below=0.4cm of Sr1, align=left, font=\scriptsize, text=gray!60!black, anchor=west, xshift=-0.5cm] 
        {\textbf{Exp. NQCC}\\-42.729(37) MHz};

    \coordinate (SrS_center) at (0.5, -0.5);
    
    \draw[bond_base] (0, -0.5) -- (2.2, -0.5); 
    \draw[bond_high] (0, -0.5) -- (2.2, -0.5);
    
    \node[atom, ball color=atom_sr] (Sr2) at (0, -0.5) {Sr};
    \node[atom, ball color=atom_s] (S1) at (2, -0.5) {S};
    
    \node[above=0.3cm of S1, text=accent_blue!80!black, font=\bfseries\footnotesize] {SrS};
    \node[below=0.4cm of Sr2, align=left, font=\scriptsize, text=gray!60!black, anchor=west, xshift=-0.5cm] 
        {\textbf{Exp. NQCC}\\-21.959(85) MHz};

    
    \node[methodbox] (theory) at (6.5, 1.0) {
        \textbf{COMPUTATIONAL INPUT}\\[0.4em]
        \scriptsize Relativistic Hamiltonian (X2Cmmf)\\
        \scriptsize CCSD(T) + Gaunt + Vib.\\
        \scriptsize \textcolor{cyan!80}{Converged Basis Set}
    };

    
    \draw[flowline] (O1.east) to[out=0, in=160] (theory.north west);
    
    \draw[flowline] (S1.east) to[out=0, in=200] (theory.south west);

    
    \node[resultbox, right=1.2cm of theory] (result) {
        \textcolor{gray}{\small New Reference Value}\\[0.2em]
        \Large $\bm{Q}(^{87}\text{Sr})$\\[0.3em]
        \huge \textcolor{alert_red}{\textbf{0.337(3) b}}
    };
    
    \draw[line width=2pt, draw=tech_dark!50, ->] (theory.east) -- (result.west);

    \begin{scope}[on background layer]
        \fill[tech_light!50, rounded corners=10pt] (-1, -2.5) rectangle (9.8, 4.5);
    \end{scope}

    
    \node[circle, fill=orange!90!yellow, text=white, font=\bfseries\tiny, align=center, inner sep=2pt, rotate=15, drop shadow] 
        at ($(result.north east)+(-0.2,0.2)$) 
        {+10\%\\vs Std.};

    \node[below=0.6cm of result, text=black!90, font=\large] 
        {$Q \propto \frac{\text{NQCC}_{\text{exp}}}{\text{EFG}_{\text{calc}}}$};

\end{tikzpicture}
\end{center}
\twocolumngrid
\bibliographystyle{aipnum4-1}
\bibliography{bibliography.bib}
\end{document}